\documentclass[11pt,notoc,hyper]{article}
\usepackage{jcappub}

\usepackage{graphicx}
\usepackage{dcolumn}
\usepackage{bm,amsmath,amssymb} 

\usepackage{epsfig,multicol,bbm} 

\def\({\left(}
\def\){\right)}
\def\[{\left[}
\def\]{\right]}

\newcommand{\LCDM}{\Lambda\text{CDM}}

\newcommand{\fsky}{f_{\rm sky}}

\newcommand{\ylmtwo}{{}_{\pm 2}Y_{lm} }

\newcommand{\ylm}{Y_{lm}}

\newcommand{\almtwo}{{}_{\pm 2}a_{lm}}
\newcommand{\almptwo}{{}_{+2}a_{lm}}
\newcommand{\almntwo}{{}_{-2}a_{lm}}

\newcommand{\almE}{a_{lm}^E}
\newcommand{\almB}{a_{lm}^B}

\newcommand{\Cl}{C_{\ell}}

\newcommand{\reffig}[1]{Figure~\ref{#1}}
\newcommand{\reftbl}[1]{Table~\ref{#1}}
\newcommand{\refsec}[1]{\S\ref{#1}}
\newcommand{\eqn}[1]{Eq.~(\ref{#1})}
\newcommand{\pp}{.}
\newcommand{\vv}{,}

\def\figdir#1{figures/#1}
\def\FIGURE#1{\begin{figure}#1\end{figure}}
\def\TABLE#1{\begin{table}#1\end{table}}

\title{Estimating the tensor-to-scalar ratio and the effect of residual foreground contamination}

\author[a]{Y.~Fantaye}
\author[b]{F.~Stivoli}
\author[c]{J.~Grain}
\author[a,d]{S.~M.~Leach}
\author[e]{M.~Tristram}   
\author[a,d]{C.~Baccigalupi}
\author[f]{R.~Stompor} 

\affiliation[a]{SISSA, Astrophysics Sector, via Bonomea 265, Trieste 34136, Italy} 
\affiliation[b]{INRIA, Laboratoire de Recherche en Informatique, Universit\'e Paris-Sud 11, B\^atiment 490, 91405 Orsay Cedex, France}

\affiliation[c]{CNRS, Institut d'Astrophysique Spatiale, Universit\'e Paris-Sud 11, B\^atiments 120-121, 91405 Orsay Cedex, France}

\affiliation[d]{INFN, Sezione di Trieste, Via Valerio 2, I-34151 Trieste, Italy} 

\affiliation[e]{CNRS, Laboratoire de l'Acc\'el\'erateur Lin\'eaire, Universit\'e Paris-Sud 11, B\^atiment 200, 91898 Orsay Cedex, France}

\affiliation[f]{CNRS,  Laboratoire Astroparticule \& Cosmologie, 10 rue A. Domon et L. Duquet, F-75205 Paris Cedex 13, France}

\emailAdd{fantaye@sissa.it}

\arxivnumber{}
\notoc

\begin{document}

\maketitle

\begin{abstract}

We consider future balloon-borne and ground-based suborbital experiments
designed to search for inflationary gravitational waves,
and investigate the impact of residual foregrounds that remain in the estimated
cosmic microwave background maps. This is achieved by
propagating foreground modelling uncertainties from the component separation, under the assumption of a spatially uniform foreground frequency scaling,
through to the power spectrum estimates, and up to measurement of 
the tensor to scalar ratio in the parameter estimation step.
We characterize the error covariance due to subtracted foregrounds,
and find it  to be subdominant compared to
instrumental noise and sample variance in our simulated data
analysis. We model the unsubtracted residual foreground contribution using a
two-parameter power law and show that marginalization over these foreground
parameters is effective in  accounting for a bias  due to excess foreground power at low $\ell$.
We conclude that, at least in the suborbital experimental setups we have simulated, foreground errors may be
 modeled and propagated up to parameter estimation  with only a slight degradation of the target
sensitivity of these experiments derived neglecting the
presence of the foregrounds. 
\end{abstract}


\section{Introduction}

All models of inflation produce tensor perturbations with an amplitude proportional to the
Hubble expansion rate during the very early phase of the universe. The
ratio of the  amplitude of tensor to scalar perturbations, denoted by $r$, is the smoking gun
for inflation and an important discriminant between classes of inflationary models ~\cite{2000cils.book.....L,2003moco.book.....D}.

The  possibility of measuring the presence of primordial gravitational waves in
the near future depends on being able to measure and characterize CMB polarization.
The decomposition of this field into two orthogonal E and B modes can  then be linked directly to
the properties of the primordial cosmological perturbations. The
E-mode polarization pattern is generated by all cosmological
perturbations while the B-mode is only generated by non-scalar sources
including gravity waves present at decoupling
\cite{Kamionkowski:1996ks,Seljak_1997,Seljak_1998}. As a result of this,
the detection of cosmological B-modes on angular scales on which lensing noise
is sub-dominant would  lend powerful observational support for 
the presence of  inflationary gravitational waves.

The B-mode signal is a small fraction of the total polarized signal
making it especially prone to contamination by systematic errors
and detector noise. Further complications arise from polarized foregrounds which
dominate  over the B-mode signal  across much of the sky~\cite{Page_etal_2007,Gold_WMAP5}, making
the detection and characterization of cosmological B-modes a very challenging
task. Any future claim of a B-mode detection, therefore, has to
rigorously show that the many potential sources of uncertainty encountered in
the CMB data analysis  methods are not biasing the result.

The upper limit on the tensor-to-scalar ratio, $r<0.28$ ($95\%$ confidence level)
\cite{Komatsu_WMAP7}, is dominated by the gravitational wave contribution to the
temperature power spectrum. Significantly improved constraints on $r$,
however, are only possible with measurements of the the B-mode
spectrum whose dominant contribution comes from the tensor perturbations.
There have been a number of studies investigating the feasibility of this
measurement in the presence of foregrounds and of the analysis techniques
that may need to be
deployed~\cite{Amarie_Hirata_lowEllPol_05,Efstathiou_lowEllPol_09,Betoule_BmodeFgPresent_2009}.

In a recent paper, \cite{Stivoli_etal_2010} combined a parametric
maximum likelihood foreground cleaning method from \cite{Stompor_etal_2009}
with a power spectrum estimation method from \cite{xpure09} in a simulated
data analysis pipeline for ground-based and balloon-borne  CMB polarization experiments. They found
that, depending on the  frequency coverage of the experiments, foreground residuals might
play a role and can potentially bias the recovery of B-modes. In cases
in which this does not happen, or external information is provided in
order to control foreground residuals, a Fisher matrix  derived estimate
indicated that a value of $r$ as low as around $0.04$ at $95\%$
confidence level might be  measured.

In this work we extend the work of \cite{Stivoli_etal_2010} in three important directions: 
\begin{enumerate}
 \item We incorporate the uncertainties of other cosmological parameters and study their effects on our
final $r$ estimates.
\item We study the effect of using a full non-diagonal  power spectrum covariance matrix on the $r$ determination.
\item  We investigate the possibility of accounting for  parameter biases from residual foregrounds 
at the power spectrum level using a simple two-parameter power law model. 
\end{enumerate}
These extensions allow us to study the impact of foreground cleaning uncertainties and parameter degeneracies on the final determination of $r$. In addition, they allow us to compare two different techniques of accounting for the unmodeled residual foreground, templates and priors at the map level and a model of the residual power spectrum at the parameter estimation level. 

Throughout this work we assume a $\LCDM$ model. 
The fiducial values of the cosmological parameters for our simulations are taken to be that of the WMAP5 best-fit
values \cite{Dunkley_WMAP5}: $ \Omega_b h^2 = 0.0227$,
$\Omega_{\rm dm} h^2 = 0.111$, $\Omega_{k} = 0$, $h = 0.719$, $\tau =
0.084$, $n_{\rm s} = 0.963$, and $A = 2.41 \times 10^{-9}$, where $\Omega_b$, $\Omega_{\rm dm}$ and $\Omega_{k}$ respectively are the density fractions of  baryons, dark matter and curvature to the critical density, a density needed for the universe to be spatially flat. $h$ is the Hubble parameter in units of 100 (km/s)/Mpc, $A$ and $n_s$ are the primordial power spectrum amplitude and spectral index parameters respectively.  For the tensor-to-scalar ratio we assume a fiducial value of $r= 0.05$
, which corresponds to a physically interesting goal that several near-term suborbital CMB polarization experiments are shooting for. Our simulated surveys are approximately optimized for measurement of this level of primordial gravity waves.

This paper is organized as follows: in Section~\ref{summary} we review
the formalism of measuring B-mode polarization and the method we use
to clean the raw CMB map from foreground contamination. In
Section~\ref{paramEst} we present our
 foreground parameter determination  methodology and  results
from applying it to simulated data in Section~\ref{result}, concluding in Section~\ref{conclusion}.


\section{CMB and foreground polarization}\label{summary}

The aim of this section is to briefly review the main assumptions of
this work, starting from the physics of CMB polarization and its
characterization in terms of `pure pseudo power spectra', then giving 
a description of our foreground simulations and cleaning technique.

\subsection{Characterization of CMB polarization}

With the onset of decoupling, scalar, vector and tensor cosmological
perturbations source a quadrupole anisotropy in the photon field. Thomson scattering in
the presence of this anisotropy then generates CMB polarization. The photons
mean free path before recombination was too short to allow for generation of 
any substantial fraction of polarized photons, while  after recombination,
there were few free electrons to act as scatterers. Hence the CMB
polarization anisotropies we see today mainly originate from the brief period before
recombination. After reionization, an additional polarized signal is
produced due to the presence of free electrons. In summary CMB
polarization depends on the physics of recombination, reionization
and the primordial perturbations, as embodied in the CAMB CMB anisotropy code~\cite{camb}.

The CMB field can be fully characterized by using the four Stokes
parameters: the intensity or temperature $T$, the $Q$ and $U$ linear
polarization vectors and the circular polarization $V$. The circular
polarization is not produced by Thomson scattering, hence only
the $T$, $Q$ and $U$ parameters are used to characterize the CMB.

The measured $Q$ and $U$ Stokes parameters vary under coordinate transformation as
headless vectors, and are related to each other by a 45 degree rotation of the
coordinate system.  The quantity $P = (Q+iU)$, however, transforms as a
spin-2 quantity under coordinate rotation i.e. $(Q \pm iU) = \exp{(\mp 2i
  \psi)}(Q \pm iU)$,  where $\psi$ is the rotation angle. The temperature
is a scalar field and can be expanded using the standard
spherical harmonics, $\ylm$. The analogous basis for a polarization tensor
field is the spin-weighted spherical harmonics basis,
$\ylmtwo$
\begin{eqnarray}
 P = \sum{\almtwo \ylmtwo} \pp
\end{eqnarray}
Instead of $\almtwo$, it is physically suggestive and convenient to
introduce their linear combinations with an even and odd parity nature
\cite{Kamionkowski:1996ks,Seljak_1998}
\begin{eqnarray}
  \almE &=& -(\almptwo + \almntwo)/2 \vv \\
   \almB &=& i(\almptwo - \almntwo)/2 \pp
\end{eqnarray}
These coefficients are named E and B modes in analogy with electric
and magnetic fields: the E-mode pattern is curl free and the B-mode
pattern is divergence free. In the absence of parity violating interactions,
the cross correlation between B and E or B and T vanishes
because B has the opposite parity of T and E. 


\subsection{Pseudo-$\Cl$ power spectrum estimation}   
The \emph{pseudo-$\Cl$} power spectrum estimation is one of the most
widely used CMB power spectrum estimators. Straightforward application
of the this technique to cut-sky polarized
CMB maps leads, however, to E-to-B leakage.
The consequence of this leakage is that
cosmologically important information contained in the CMB B-modes is
overwhelmed by the statistical uncertainty of the (much larger)
E-modes.

The \emph{pure-pseudo} power spectrum estimators, adopted in this work, define a special
conditions on the boundary of the cut-sky region in such a way that
only the pure E and B-modes are retained 
\cite{Bunn_EB_2003,Smith_Zald_2007,Smith_2006,xpure09}. We used the {\sc
  Xpure}\footnote{\texttt{www.apc.univ-paris7.fr/$\sim$radek/pureS2HAT.html}}
code~\cite{xpure09}, which is a generalization of the {\sc Xspect}
and {\sc Xpol} codes~\cite{Xspect_2005}, for estimating the pure
pseudo-power spectra, $C^{EE}$ and $C^{BB}$. {\sc Xpure} uses the {\sc
  S$^2$HAT} library\footnote{\texttt{www.apc.univ-paris7.fr/$\sim$radek/s2hat.html}} \cite{s2hat_gpu}
- an efficient parallel implementation of spin-weighted spherical
harmonic transforms.


\subsection{Foreground modelling and cleaning} \label{foreground}
In this section we summarize the foreground cleaning of our data analysis pipeline. Further
details may be found in \cite{Stivoli_etal_2010},
\cite{Stompor_etal_2009}.  Our focus is on
typical configurations and sensitivity levels of future ground-based
and balloon-borne observations, which is summarized in \reftbl{suborbExp_param},
and correspond roughly to the nominal expected performances of
POLARBEAR \cite{polarbear} and EBEX \cite{2010SPIE.7741E..37R} experiments. Our analysis method should also be of interest to other experiments like SPIDER \cite{Odea_SPIDER} whose frequency bands roughly overlap with those of POLARBEAR.

\TABLE{
\centering
$\begin{array}{@{\hspace{+0.0in}}c} 
\begin{tabular}{|c|c|c|c|} 
\hline
Case & \bf{\emph{$\fsky$}} & {Frequency [GHz]} & {Noise level [$\mu K_{\rm CMB}$.{\rm arcmin}]} \\ \hline 
Balloon-borne & $1\%$ & 150, 250, 410 & 5.25,   14.0,  140.0 \\ \hline 
Ground-based & $2.5\%$ & 90, 150, 220 & 10.5,  10.5,   31.5\\ \hline 
\end{tabular} 
\end{array}$ 
\caption[survey parameter values]{Balloon-borne and ground-based
  experimental configurations. All channels have a Gaussian beam width
  of FWHM$=8'$. \label{suborbExp_param}}
}

According to previous studies of the intensity of thermal dust emission, the area centered around
RA $=62^o$ and DEC $= -45^o$ is favorable in terms of
foreground brightness and contrast as well as being accessible by suborbital experiments.
This area, therefore, has been the target for
many past CMB suborbital experiments i.e.  Boomerang \cite{Boomrang},
BICEP \cite{bicep}, SPT \cite{spt}, QUAD \cite{quad_2009}, ACBAR
\cite{acbar} and QUIET~\cite{2010arXiv1012.3191Q}. It is also an
area of choice for current and future ground-based and balloon-borne
experiments, including POLARBEAR and EBEX respectively.

We consider the frequency interval from 90--410 GHz (see
\reftbl{suborbExp_param}) and so our Galactic sky model contains two diffuse
polarized foreground components, namely thermal dust and synchrotron. The total
intensity simulation is based on observations by \cite{Haslam_1982}
and modelling of the frequency scalings by \cite{Finkbeiner_1999}.
The polarization signal was then modelled as a $10\%$ fraction of the
total intensity with the pattern of the polarization angles on large
angular scales given by the WMAP dust polarization template \cite{Page_etal_2007}. On
smaller angular scales we added a Gaussian fluctuation power adopting a recipe first presented in \cite{Giardino_2002}. We refer to \cite{Stivoli_etal_2010} for a full description of the model.

As shown in \cite{Page_etal_2007} polarization foreground power is
comparable or higher than the CMB B-mode power on all scales which are
accessible to the WMAP experiment. Therefore, a robust foreground
cleaning procedure needs to be applied to the data prior to B-mode
measurement. 

Our approach to this problem is to use the parametric component
separation algorithm proposed by
\cite{Stompor_etal_2009} and applied in \cite{Stivoli_etal_2010} for
CMB suborbital polarization experiments. This algorithm solves the
component separation problem in two steps. Firstly, the maximum likelihood
values of the free parameters of the foreground frequency scalings are determined
using a spectral index `profile likelihood' that has been analytically derived under the
assumption of constant foreground scaling across the entire map. Secondly,
having fixed the foreground frequency scalings, the component amplitudes are estimated in
a pixel by pixel, using a generalized least-squares algorithm.

Specifically the spectral index likelihood is given by
\begin{equation}\label{eqn:slopelikeMpix}
-2\ln {\cal L}_{spec}\(\bm{\beta}\)  = \hbox{{\sc const}} -
\( \bm{A}^t\, \bm{N}^{-1}\,\bm{d}\)^t\,\(\bm{A}^t\,\bm{N}^{-1}\,
\bm{A}\)^{-1} \( \bm{A}^t\,\bm{N}^{-1}\,\bm{d}\) \vv
\end{equation}
where $\bm{d}$ is the data, $\bm{A} \equiv\bm{A}  \left(\bm{\beta}\right)$ 
is a component `mixing' or frequency
scaling matrix with a total of $N_{spec}$ free `spectral parameters', 
$\bm{\beta}$,  which are to be estimated, and $\bm{N}$ is the 
noise covariance matrix of the data. The maximum 
likelihood values of  $\bm{\beta}$ are then substituted into
 \begin{eqnarray}
\bm{s} &=& \(\bm{A}^t\,\bm{N}^{-1}\,\bm{A}\)^{-1}\,\bm{A}^t\,\bm{N}^{-1}\,\bm{d} \vv
\label{eqn:step2amps} \\
\bm{N_s} &\equiv& \(\bm{A}^t\,\bm{N}^{-1}\,\bm{A}\)^{-1} \vv
\label{eqn:noiseCorrOptim}
\end{eqnarray}
to solve for the amplitudes, $\bm{s}$, of the different components and their
covariance $\bm{N_s}$. This algorithm is implemented with a Markov chain 
Monte Carlo based maximization in the {\sc
  miramare}\footnote{\texttt{people.sissa.it/$\sim$leach/miramare}}
code.
This formalism, like other likelihood based component separation methods, 
has an additional advantage of easily parametrizing and propagating 
instrumental features like calibration uncertainties to the final 
cleaned maps and covariance matrices.

As we have seen above, the typical parameters to be estimated in the
component separation phase are the foreground and CMB amplitudes, and
foreground spectral parameters. Given the three frequency channel
designs of our balloon-borne and ground-based setups, we can only deal with
one dominant foreground, which corresponds to the thermal dust for
both experiments' frequency coverage.  In the analysis by \cite{Stivoli_etal_2010}
it was shown that the contamination of the unmodeled and hence uncleaned
synchrotron component in the final CMB map is negligible
for the balloon-borne experiment, owing to the choice of frequency bands.
On the other hand, for the ground-based case,
which has a lower frequency band at 90 GHz where the synchrotron emission may
still be relevant, the synchrotron contamination is high enough to bias the dust
spectral index estimate thereby compromising the cosmological B-mode signal.
Either a prior on the dust spectral index or external information on the synchrotron is
therefore required for restoring a cleaning performance comparable to the balloon-borne case.

Although we assume a constant spectral index across the observed region, 
the likelihood \eqn{eqn:slopelikeMpix} can accommodate a spatially varying spectral index, 
for instance by evaluating the likelihood on separate low resolution pixels 
\cite{Eriksen_2006} or by modeling spectral index variation as a 
new component \cite{Stolyarov_2005}. Both of these foreground model 
extensions come at the cost of introducing of extra map-level foreground 
free parameters and ideally would benefit from the luxury of having 
extra data channels. Conversely, the approach assumed here be thought 
of as a potentially useful ``minimally parametric" approach that may 
be an appropriate  for the partial sky, low foreground contrast 
setting we are investigating. As argued by other authors \cite{Tucci_2005, Verde_2006} 
any unmodeled spectral index variations might in the end manifest 
themselves as a residual foreground contamination with a powerlaw 
$\Cl$ spectrum - a theme that we will return to in \refsec{rfg_section}.

The main aim of this paper is to demonstrate that instead of
relying on external priors for the dust spectral scalings or templates
for the synchrotron component, one can perform a self-contained
component separation and then account for diffuse foreground bias in the
parameter estimation phase through a simple two-parameter residual foreground power spectrum model.

\section{Parameter estimation with foreground cleaned CMB templates} \label{paramEst}
In this section we discuss the parameter estimation methodology that we have developed for the purposes
of deriving unbiased constraints on the cosmological parameters in the presence of foregrounds. Our
methodology incorporates the full power spectrum covariance matrix and a model for the residual foreground power spectrum. 

\subsection{Likelihood}

The likelihood of the data, compressed into power spectrum bands, given a model is assumed to have an offset lognormal distribution \cite{Bond_logNormal}  
\begin{equation}\label{logNL}
 -2\ln{\mathcal{L}} = 
\sum_{\ell \ell'}[\hat{z}_{\ell}^{XX}-z_{\ell}^{XX}]^t\mathcal{Q}_{\ell \ell'}
 [\hat{z}_{\ell'}^{XX}-z_{\ell'}^{XX}]  \vv
\end{equation}
where $\hat{z}_\ell = \ln(\hat{\Cl})$, $z_\ell = \ln(\Cl)$, and $\mathcal{Q}_{\ell \ell'} = \hat{\Cl}\Sigma^{-1}_{\ell \ell'}{\hat C_{\ell'}}$ is the 
local transformation of the inverse covariance matrix, 
$\Sigma^{-1}_{\ell \ell'}$, to the lognormal variables 
$z_\ell$. Here $\hat{C}_{\ell}$ and $C_{\ell}$ are the 
estimated and model power spectra respectively, 
and $XX$ denotes the EE and BB spectra. 

For the balloon-borne and ground-based experiments considered here, we
used only the EE and BB power spectra generated using the {\sc
  Xpure} pipeline \cite{xpure09}. The minimum and maximum multipoles
considered are $\ell_{\rm min} = 20$, $\ell_{\rm max} = 1020$. To
reduce cut-sky effects, we averaged the spectra into multipole bins of width $\Delta \ell = 40$.

To explore the parameter space of the likelihood, we used the 
{\sc CosmoMC} sampler \cite{cosmomc} in its
multi-chain mode. The convergence of the chains was verified by
demanding that the `Gelman-Rubin statistic' $R<0.1$, which measures 
 the variance of the chain means divided by
 the mean of the chain variances.

\subsection{Planck prior} \label{priors}

The suborbital experiments we are considering here are specifically
designed to measure the EE and BB spectra, which are particularly sensitive to
the reionization optical depth, $\tau$, and the
tensor to scalar ratio, $r$. However, these experiments are not optimized to make CMB
temperature measurements and lack the polarization information on the largest angular scales. We therefore
imposed a Planck prior to give parameter constraints that are representative of the
situation expected when these experiments begin taking or analyzing data.  More specifically, the joint analysis accounts for the fact that the constraints on $r$ are expected to be correlated with the other
cosmological parameters, particularly $n_{\rm s}$.

Planck \cite{Planck_mission_2011}, which is currently observing from the L2 Lagrangian point, is
expected to obtain improved constraints compared to WMAP by a factor
of a few on the main cosmological parameters. Our 
simplified Planck TT, TE and EE polarization power spectra, 
are simulated with a modified {\sc futurCMB} code
\cite{Perotto_FutureCMB_06} using the HFI 100, 143, 217
GHz channels and with specifications given in \cite{Planck_mission_2011}.  For comparison we also investigated a mock WMAP likelihood based on the noise levels of the WMAP Q, V, W bands in temperature and V and W bands in polarization \cite{Jarosik_WMAP7}. For these likelihoods we used the exact expression from \cite{Lewis_LensedCMB_05}.

We excluded the satellite BB power spectra from our analysis  as our main interest in this work
is to investigate the constraining power of future suborbital B-mode experiments and to develop
an appropriate analysis technique. It is likely that a measurement of the B-mode from reionization
at low $\ell$ with Planck will require a more specialized treatment due to the higher foreground
to signal level \cite{Efstathiou_lowEllPol_09,Katayama_Komatsu_2011,Bonaldi-Ricciardi_2011,Armitage-Caplan}.

\subsection{Covariance matrix}

\FIGURE{
$\begin{array}{lll}
\includegraphics[width=3in,height=2in]{\figdir{balloon_log_covMat_EE.eps}} &
\includegraphics[width=3in,height=2in]{\figdir{balloon_log_covMat_BB.eps}} \\
\includegraphics[width=3in,height=2in]{\figdir{ground_log_covMat_EE.eps}} &
\includegraphics[width=3in,height=2in]{\figdir{ground_log_covMat_BB.eps}} 
\end{array}$
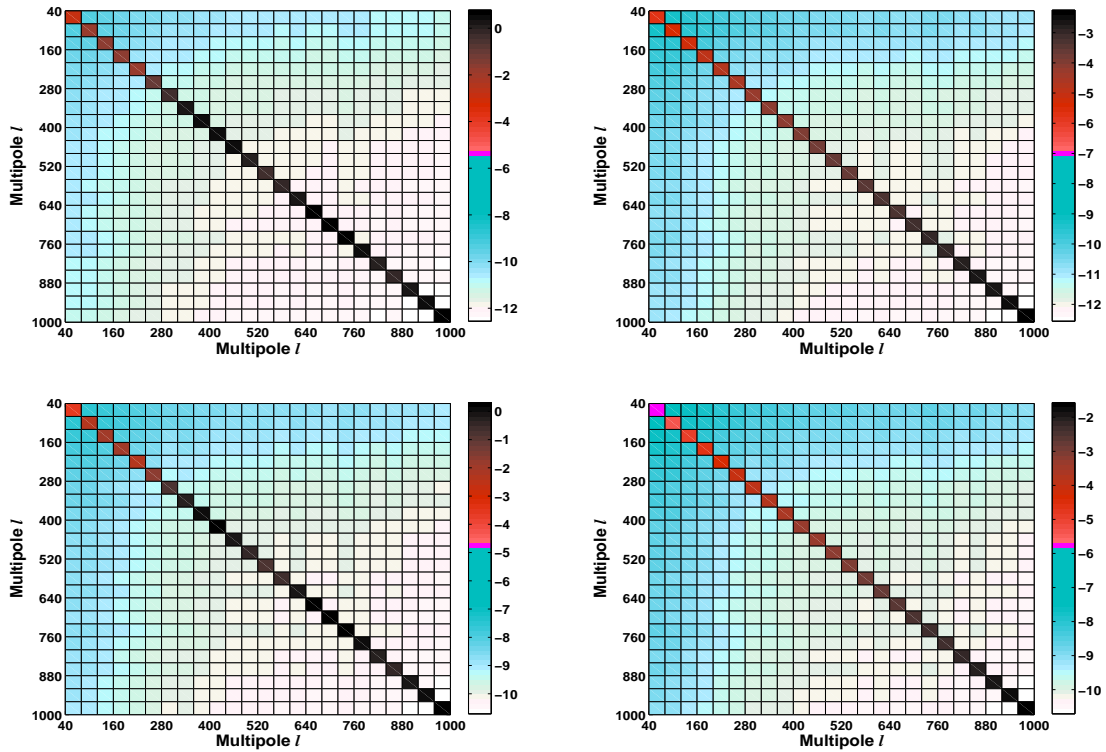
\caption{The CMB EE (left) and BB (right) power spectrum covariance
  matrix for the balloon-borne (top panels) and ground-based (bottom
  panels) experiments in binned-multipole space. The colour scale
  is logarithmic in units of $\mu$K$^2$. The different  terms  
  that specify the full   covariance matrix are given in 
  Eq.~(\ref{eqn:sigma_fisher}).  The off-diagonal elements, which are 
  are several orders of magnitude smaller than the diagonal parts, 
  are caused by the presence of correlated residual foregrounds.}
 \label{fig:balloon-covmat}
}

The form of the pseudo-$C_{\ell}$ covariance matrix   
in the presence of an instrumental white noise and a residual foreground, as
defined in Eq. (20) of \cite{Stivoli_etal_2010}, is given by
\begin{equation}\label{eqn:sigma_fisher}
{\Sigma}_{\ell \ell'}  \equiv  \frac{2(C_\ell^{\rm CMB} + N_\ell^{\rm noise})^2}{(2\ell+1)\fsky}
 + \frac{4(C_\ell^{\rm CMB} + 
N_\ell^{\rm noise})C^{\rm rfg}_{\ell}}{(2\ell+1)} + 2C^{\rm rfg}_{\ell} C^{\rm rfg}_{\ell'} \vv
\end{equation}
where the first term in this equation represents the sample plus noise variance of the
estimated pure B-mode spectra, which are assumed to be
diagonal in multipole bin-space. 
In our analysis, this term is computed via 500 Monte-Carlo simulations.
The second term corresponds to the covariance 
of the cross correlation between the pure CMB+noise and the 
residual foreground power. The $C^{\rm rfg}_{\ell}$ is estimated using 
 the model given by Eq. (27) of \cite{Stivoli_etal_2010},  which is where the estimated foreground power spectrum and the uncertainty on the estimated foreground spectral indices, $\Delta\beta$, are both factored in. The last term in \eqn{eqn:sigma_fisher} 
represents the covariance due to the residual foreground, 
and is fully correlated between different bins. 
This term corresponds roughly to the tensor-product
term found in equation~(A4) of~\cite{Stompor_etal_2009}, which indeed
describes the correction to the noise correlation matrix,
Eq.~(\ref{eqn:noiseCorrOptim}), due to the foreground residuals. 
Note that this covariance formula in \eqn{eqn:sigma_fisher} corrects and supersedes 
the one in Eq. (31) of \cite{Stivoli_etal_2010}.
 
In \reffig{fig:balloon-covmat} we show the full covariance
matrix computed for the `balloon-borne basic' and `ground-based template' setups, which are two
foreground clean cases investigated by~\cite{Stivoli_etal_2010}, and discussed further in the next section.
The dominant contribution to these covariances are from the variance of the
CMB+noise spectra, given by the first term in Eq.~(\ref{eqn:sigma_fisher}). 
It is clear that the off-diagonal elements are at least one order of magnitude smaller than the diagonal elements, reflecting the fact that the amount of the residual
foregrounds for these two setups is subdominant.  The overall amplitude of the foreground covariance
is controlled by a factor $\Delta\beta^2$ and the decay in correlations towards
high $\ell$ reflects the decaying power spectrum of the residual foregrounds.

\subsection{Modelling the residual foreground}\label{rfg_section}

Foreground residuals in the recovered CMB map resemble the underlying foreground emission
and bias the estimated CMB signal. If we could reliably account all foregrounds in the component separation phase, the
resulting  effect of the residuals on the estimated CMB signal could  straightforwardly be propagated
in the covariance matrix using \eqn{eqn:sigma_fisher}. For the type of experiments we consider here, however,
this is not the case since we have insufficient frequency coverage to deal with  the synchrotron component
which hence requires external information in the form of a strong prior on the foreground spectral indices or an external template.

The use of templates or external priors on the dust spectral
parameters has its own limitation as we do not have polarization foreground  surveys
at frequencies, sensitivities  and angular scales of interest. Furthermore it is
commonly through extrapolation of low frequency observations that
the required template is produced, but this process is prone to astrophysical complications. 
Ideally we want to perform a self contained analysis.

Modelling of sources of residual foreground power in temperature at high $\ell$ is of course quite common
in the literature~\cite{Hinshaw_2003,Spergel_2007,spt}. We propose
that a low $\ell$ excess of polarized power due to residual foregrounds may be modelled in a similar way using a phenomenologically motivated two-parameter power law model. This  parametrization is inspired by WMAP modelling \cite{Page_etal_2007,Gold_WMAP5} and  the fact that the dominant contribution to the residual foreground is due to the unmodeled synchrotron component, it is given by
\begin{eqnarray}\label{rfg_model}
  D_{\ell}^{\rm rfg} = A_{\rm rfg}\hat{D}^{BB}_{\ell_0}\(\frac{\ell}{\ell_0}\)^{\beta_{\rm rfg}},
\end{eqnarray}
where $D_l = \ell(\ell+1) \Cl/2\pi$, $\ell_0$ is the pivot multipole where the residual model is
normalized at and $\hat{C}^{BB}_{\ell_0}$ is the  B-mode power spectrum estimate in the lowest bin $\ell_0$, taken here as a normalization constant. $A_{\rm rfg}$ is the fractional value of the residual
amount relative to $\hat{C}^{BB}$ at ${\ell_0}$, and $\beta_{\rm rfg}$ is the residual spectral index parameter. We fit these model parameters together with other cosmological parameters in the parameter estimation phase.  We normalize $D_\ell^{\rm rfg}$ at $\ell_0 = 40$, the central value of the first multipole band in our data. 


\section{Results and discussion}\label{result}

In this section we present our findings  concerning the effect of marginalizing over residual diffuse foregrounds at the parameter estimation step, and  concerning the effect of incorporating a full non-diagonal
 covariance matrix on recovered parameter values. 

\FIGURE{
\centering
\includegraphics[width=5.5in,height=4.5in]{\figdir{cl_vs_residuals.eps}}
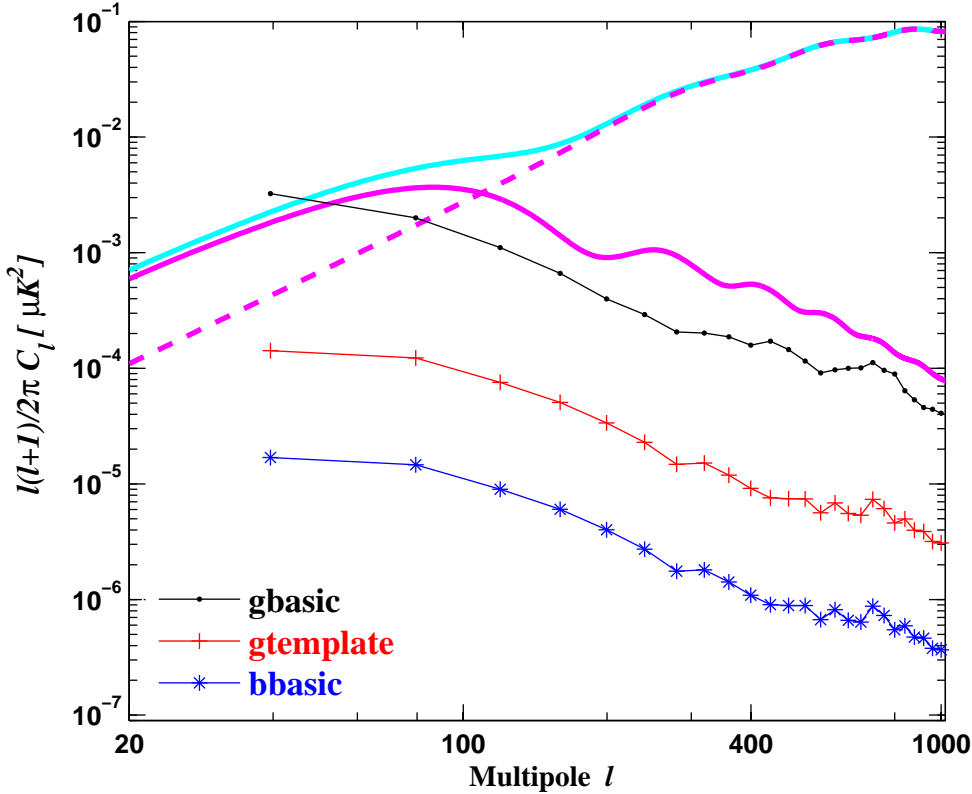
\caption{ B-mode power spectra of the foreground residuals in the CMB map. The `\emph{bbasic}' (starred blue curve) and `\emph{gtemplate}' (crossed red curve) cases are for the balloon-borne and ground-based (with external template) experiments in which successful foreground cleaning is achieved. The `\emph{gbasic}' (dotted black curve) case represents a ground-based experiment in which
a high level of residual synchrotron remains, due to insufficient frequency coverage, leading to parameter biases
in $r$ if left unmodeled. For comparison we show the model B-mode spectra from tensor modes  for $r=0.05$ (solid magenta curve),  lensing (dash-dash magenta curve) and the sum of the two (sold cyan curve)}
 \label{fig:cl_vs_residuals}
}

We considered  three cases of map-level component separation.  The first  two, which are  representative of
self-contained component separation  attempts,  includes the `\emph{bbasic}' (balloon-borne
basic) and `\emph{gbasic}' (ground-based basic)  cases. The
 third case,  in which  the ground-based experiment relies on an external  low frequency template
to remove the synchrotron component, is referred to as `\emph{gtemplate}'
(ground-based case with template cleaning). In \reffig{fig:cl_vs_residuals} 
we compare the input B-mode power spectrum with the level of residual foregrounds 
left  in the  estimated B-mode  power spectrum from these three setups.

\FIGURE{
\centering
\includegraphics[width=5in,height=6in]{\figdir{bbasic_gtemplate_gbasic_planck_priorTest.eps}}
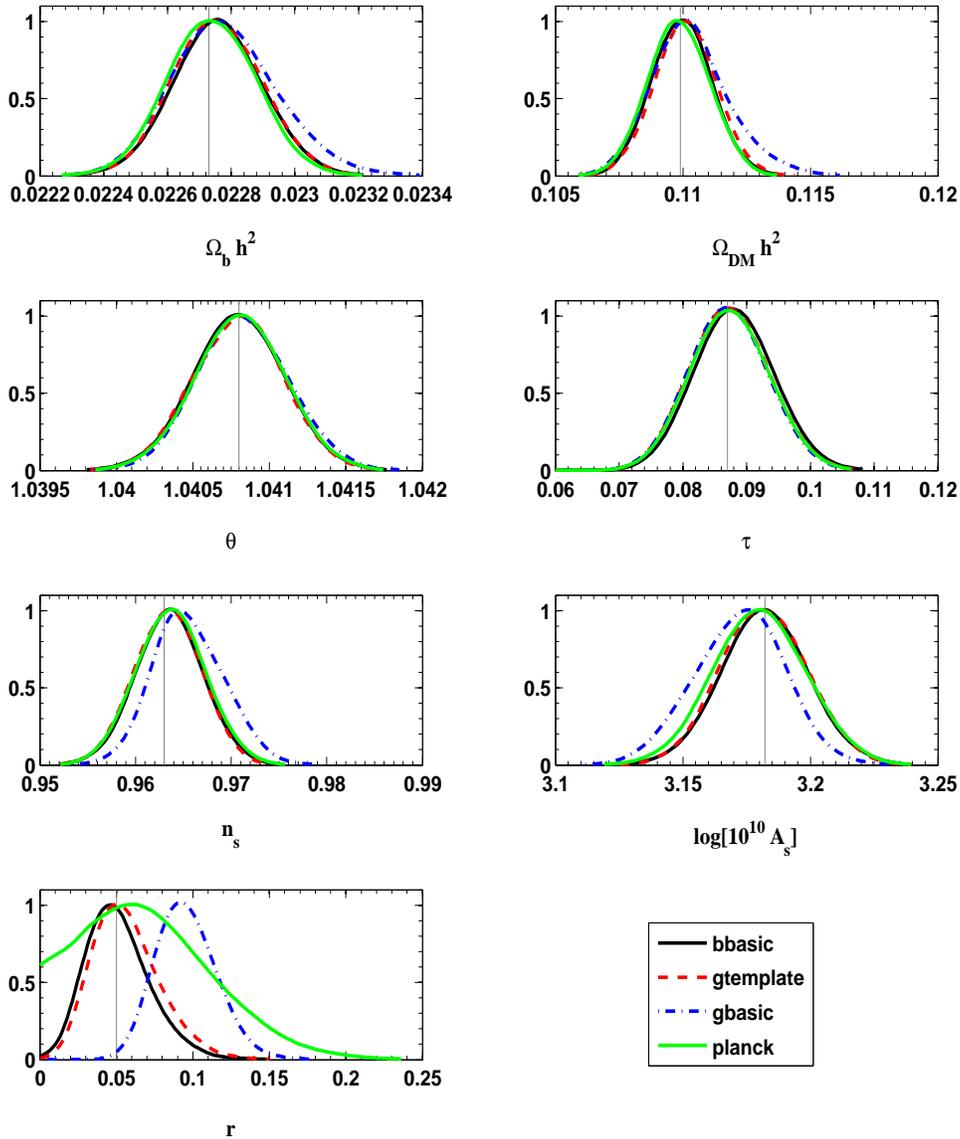
\caption{ Likelihoods for the seven cosmological parameters from the ground-based and
balloon-borne experiments. As in Figure~\ref{fig:cl_vs_residuals}, the \emph{bbasic} and
\emph{gtemplate} refer to a balloon-borne and a ground-based experiments (with external template) where successful foreground cleaning has been applied. The \emph{gbasic} case shows a parameter bias in the value of $r$ due to foreground residuals.
The vertical lines correspond to the the input values of the parameters.}
 \label{fig:suborb_nofg}
}

\begin{figure}
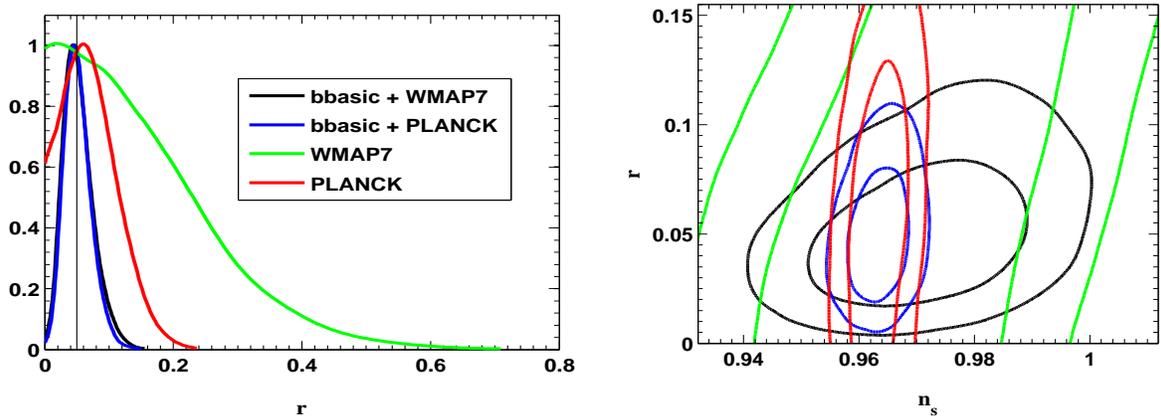

\centering
$\begin{array}{cc}
\includegraphics[width=3in,height=2.2in]{\figdir{bbasic_priorTest_p12.eps}} &
\includegraphics[width=3in,height=2.2in]{\figdir{bbasic_priorTest_p8_vs_p12.eps}}
\end{array}$
\caption{Effect of priors : imposing Planck like or WMAP like priors on the cosmological parameters which are not well measured by the suborbital experiments has negligible effect on the constraint of $r$. This is because the $r$ constraint is mainly dominated by suborbital B-mode data.}
 \label{fig:effect_prior}
\end{figure}

We show in \reffig{fig:suborb_nofg} the values of the  estimated parameters for
the above three cases, \emph{bbasic}, \emph{gbasic} and \emph{gtemplate},  as well as the Planck only case for comparison. As can be seen from this figure,  the constraint on $r$ is dominated by the suborbital experiments, while the constraints on the other parameters are dominated by the Planck prior. This is further illustrated in \reffig{fig:effect_prior} where we compare the $r$ constraint from \emph{bbasic} joint analysis and from the Planck and WMAP priors alone. The right panel of this figure shows that constraining power of the Planck prior suppresses the $n_{\rm s}-r$ degeneracy, which explicitly validates the widely-used single-parameter Fisher matrix $r$ forecasts, also assumed by \cite{Stivoli_etal_2010}.

 Returning to \reffig{fig:suborb_nofg} it is clear that  for both the \emph{bbasic} and \emph{gtemplate} cases, the input values of all parameters are recovered without bias, while for the \emph{gbasic}
case the  estimated $r$ value is biased  positive by more than
$3\sigma$. This is because for the balloon-borne or the ground-based with template
cleaning setups the unaccounted foreground residual is  negligible, while for the
 attempt at a self-contained ground-based analysis  case,
\emph{gbasic}, the foreground residual is significant, and hence
adds power to the B-mode  power spectrum estimates on  large angular scales. 
This extra  large-scale power then  biases the $r$ value estimate, as shown \reffig{cl_vis}.

\subsection{Foreground control at the level of parameter estimation}
The effect of introducing a residual foreground model Eq.~\ref{rfg_model} in our parameter estimation
phase is clearly shown in \reffig{fig:suborb_2pfg} where the
previously found bias on $r$ is now eliminated and the input values of all
the cosmological parameters are recovered with the same precision as
in the balloon-borne and template cases. 
In the same plot we show also the \emph{gbasic} and \emph{bbasic} cases 
for comparison. The constraints on the residual model parameters are,
however, poor.

\FIGURE{
 \includegraphics[width=6in,height=2in]{\figdir{gbasic_1D_2paramFg_wplanck.eps}}
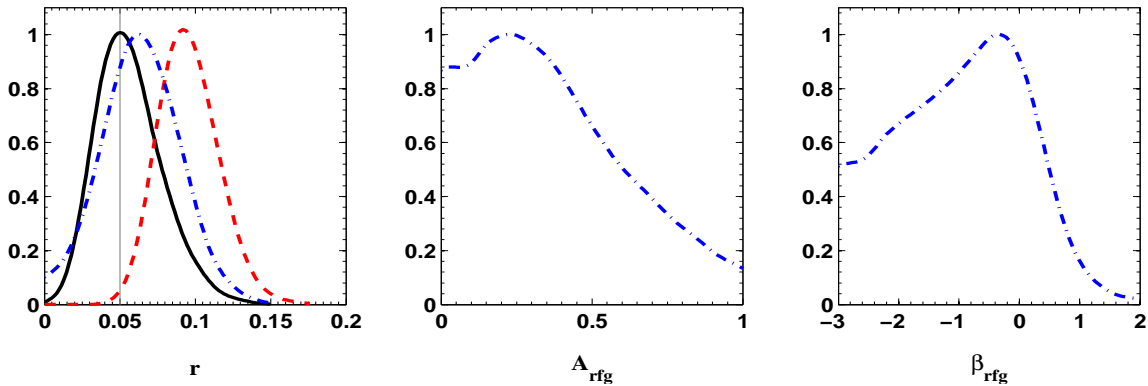
\caption{Likelihoods for the ground-based cases:  the solid black curve is for the \emph{gtemplate},
 the dash-dash red curve is for the  \emph{gbasic} and the dot-dash blue curve is for \emph{gbasic}
 with the residual foreground model given by \eqn{rfg_model} included in the
  parameter estimation.   Marginalizing over the residual foreground
  model parameters debiases the \emph{gbasic} $r$ estimate, $r=0.096\pm0.019$, towards the input value of 0.05, obtaining $r=0.064\pm0.025$, which can be compared with $r=0.057\pm0.021$ in the \emph{gtemplate} case.} 
 \label{fig:suborb_2pfg}
}
\FIGURE{
 \includegraphics[width=6in,height=2in]{\figdir{gbasic_2D_2paramFg_wplanck.eps}}
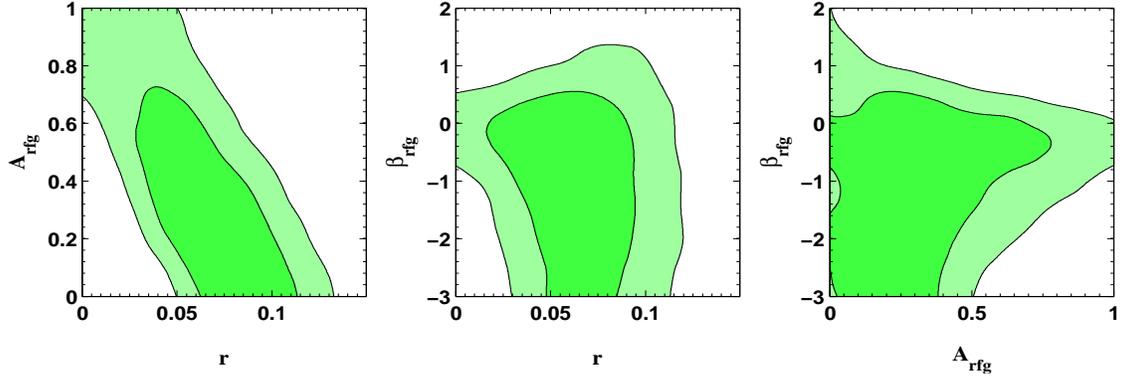
\caption{68\% ( dark green) and 95\% (light green) confidence intervals 
of the likelihood contours illustrating degeneracies among the 
tensor-to-scalar ratio, $r$, residual foreground amplitude,
$A_{\rm rfg}$, and spectral index, $\beta_{\rm rfg}$, parameters, for the \emph{gbasic} case
with the residual foreground accounted at the parameter estimation level.
This illustrates how adding foreground power, $A_{\rm rfg}$, to the fit debiases the $r$ estimate towards the input value of 0.05, and how the uncertainty $\beta_{\rm rfg}$ degrades the statistical significance of the $r$ detection.}
 \label{gbasic_wrfg_2D}
}

\FIGURE{
\centering
\includegraphics[width=5in,height=4in]{\figdir{gbasic_visualize_2paramFg.eps}}
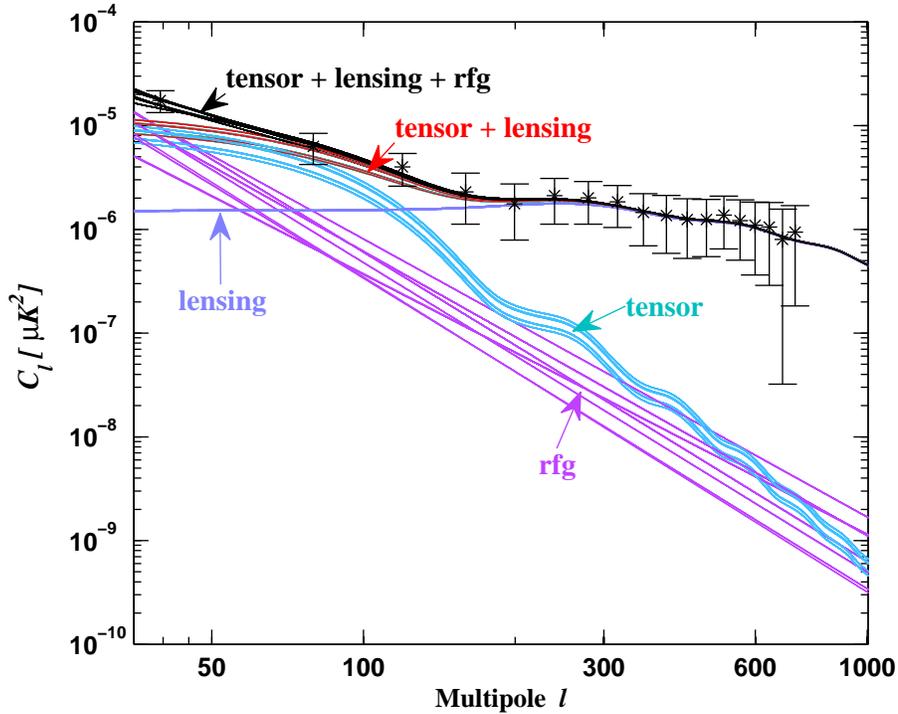
\caption{Illustration of the model $\Cl$ together with the simulated
  data for the ground-based experiment with basic foreground setup, \emph{gbasic}. The solid lines are the different  physical components that  make up 
  the total model $\Cl$.  We found the parameters that maximize the likelihood using an
  MCMC search:  the model $\Cl$ curves shown here are selected randomly from the MCMC chain after initial burn-in. This figure demonstrates how a residual foreground model can be used to subtract away excess foreground   power at low $\ell$.}
 \label{cl_vis}
}

To understand the effect of incorporating the residual foreground
model in our parameter estimation, we show in \reffig{gbasic_wrfg_2D}
the two dimensional contours of the likelihood. The left panel
 shows the anti-correlation between $r$ and the residual foreground
amplitude, $A_{\rm rfg}$. The amplitude parameter reflects the ratio
of the  residual foreground signal to the total $\Cl$ at the pivot multipole
$\ell_0=40$. From this degeneracy plot we can clearly see the reason why
the  estimated $r$ value from the \emph{gbasic} is biased:  The presence of
 residual foregrounds at low $\ell$ can mimic a positive tensor-to-scalar
ratio. The middle and right panels in \reffig{gbasic_wrfg_2D} show the
degeneracy between $r$ and $\beta_{\rm rfg}$  and $A_{\rm rfg}$ and
$\beta_{\rm rfg}$ respectively. In both these cases the degeneracies are 
weak, while the residual spectral index parameter, $\beta_{\rm rfg}$, is poorly
constrained. It is therefore safe to assume that despite the poor
knowledge of $\beta_{\rm rfg}$, our power law residual foreground model
can disentangle the residual from the CMB signal and lead to a similar
precision measurement of $r$ from  the self-contained analysis of  our simulated ground-based
experiment.

To  show the above points more clearly, we plot in \reffig{cl_vis}
the data together with the model $\Cl$ curves. This  ensemble of curves
corresponds to  models drawn from a converged MCMC
chain. By comparing the `tensor' BB curve with the residual
foreground, `rfg',  curve, we see that the $\Cl$ region that is  most
affected by the residual foreground component coincides  with the region where most
of the gravity wave signal is present, at $\ell<200$.  This plot also illustrates the importance of
measuring the B-mode power spectrum at $l>200$ in order to be able to estimate and subtract off the B-mode spectrum
from lensing.

\subsection{Effect of off-diagonal terms in the covariance matrix}

Up to this point we were using a covariance matrix in which small
off-diagonal components from the modelled residual foregrounds are
ignored.  In this  section  we address the question of how
incorporating such off-diagonal terms in the covariance matrix affect
our results.

\FIGURE{
\centering
\includegraphics[width=5in,height=6in]{\figdir{gtemplate_bbasic_covmat_effect.eps}}
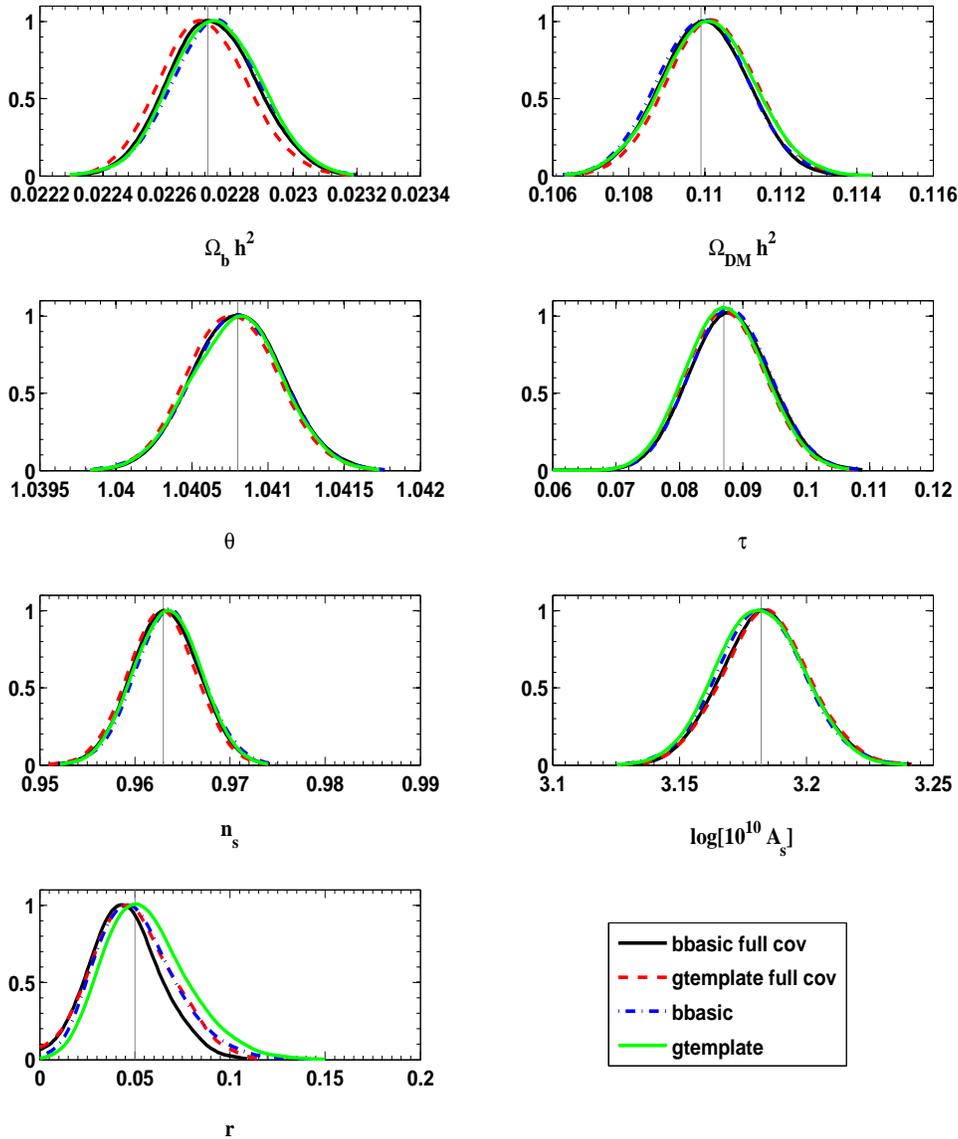
\caption{Effect of ignoring off-diagonal elements in the power spectrum covariance matrix on the
  constraints of the cosmological parameters. Plots illustrating the full
  covariance matrices for both the balloon-borne and ground-based experiments are
  shown in \reffig{fig:balloon-covmat}. Runs with `full cov` in the legend incorporates a full covariance matrix, while otherwise a diagonal covariance matrix is used. }
 \label{fig:param_wcovmat}
}

The expression that combines all propagated uncertainties starting from
component separation to power spectrum estimation is given by
\eqn{eqn:sigma_fisher}. The full covariance matrix for the \emph{bbasic} and \emph{gtemplate} cases is shown in
\reffig{fig:balloon-covmat}. We note that off-diagonal terms in both cases are
 several orders of magnitudes  smaller than the diagonal  elements.

In \reffig{fig:param_wcovmat} we compare the recovered values of our
seven base parameters using diagonal (\emph{bbasic} and \emph{gtemplate}) and
full covariance matrix (\emph{bbasic full cov} and \emph{gtemplate
full cov}).  For these cases, the effect of ignoring off-diagonal terms is  small,
 with the shift on $r$ being slightly bigger in the \emph{gtemplate} case.  This
is because, as shown in \reffig{fig:balloon-covmat}, the residual  foreground level
is higher for this case. We conclude that at least for  these cases, where the off-diagonal
terms are small as compared to the diagonal ones, ignoring the off-diagonal terms has
little impact on our results.

\section{Conclusions}\label{conclusion}
Foreground cleaning and the propagation of uncertainties related to this process 
is one of the outstanding issues in CMB data analysis for ongoing and future 
experiments. This is particularly  important for CMB polarization measurements and 
the  search for B-modes, for which foregrounds are expected to 
be sizeable at all frequencies and across most of the sky. 

Based on the foreground cleaning procedure presented in \cite{Stompor_etal_2009}, 
we developed and completed recent results from \cite{Stivoli_etal_2010} 
concerning the propagation of foreground cleaning uncertainties in the main 
steps of CMB data analysis,  with a focus on measuring the tensor to scalar ratio 
of cosmological perturbations. We considered typical experimental setups for proposed 
ground-based and balloon-borne experiments,  for which the foreground contrast might
be weak enough to justify the assumption of a spatially constant spectral index built into our parametric method.

These experiments  operate at frequencies between 90 and 410 GHz, 
where the main diffuse foreground components  are polarized synchrotron and thermal dust 
emission. We examined configurations in which a significant  source of bias remains after
the foreground cleaning phase,  due to residual synchrotron 
emission from lower frequencies. In this case, we  introduced a method for
correcting the data for the presence of 
such a residual, simply  by parametrizing the residual foreground power  at low $\ell$, 
 and then marginalizing over its  amplitude and powerlaw index. We  found that this technique   debiases the $r$ measurement, although in the case we studied the foreground parameters are poorly constrained, degrading the $r$ detection significance.

We characterize the error covariance due to subtracted foregrounds,
and find it  to be subdominant compared to
instrumental noise and sample variance in our simulated data
analysis.

We conclude that for  the experimental configurations and   template-based simulated 
foreground signals investigated here, it is possible to propagate foreground
subtraction uncertainties and marginalize over foreground residuals, without biasing
the final parameter determination results.  Therefore, the  expected performance
of these  multi-frequency experiments,  usually quoted in the absence of a properly
simulated data analysis procedure,  might only be marginally  degraded by the presence of
foreground uncertainties and residuals.

\section*{Acknowledgments}
YF thanks APC for hospitality while part of this project was completed. His work there was supported in part by French National Research Agency (ANR) through COSINUS program (project MIDAS no. ANR-09-COSI-009).
Some of the results in this paper have been derived using the HEALPix
\cite{healpix} package. We acknowledge the use of {\sc CAMB}~\cite{camb}
and {\sc CosmoMC} \cite{cosmomc} packages.
\newpage
\newpage

\bibliography{rfg}

\end{document}